\pdfoutput=1

\documentclass[11pt]{article}

\usepackage[final]{acl}

\usepackage{times}
\usepackage{latexsym}

\usepackage[T1]{fontenc}

\usepackage[utf8]{inputenc}

\usepackage{microtype}

\usepackage{inconsolata}

\usepackage{booktabs}
\usepackage{graphicx} 
\usepackage{graphics}
\usepackage{amssymb}
\usepackage{amsmath}
\usepackage{bbm}
\usepackage{pifont}
\usepackage{makecell}
\usepackage{resizegather}
\usepackage{etoolbox}
\usepackage{booktabs,makecell,tabularx} 
\usepackage{enumitem}
\usepackage{mathtools}
\usepackage{cleveref}
\usepackage{multirow}
\usepackage{booktabs,array}
\usepackage{color, colortbl}
\usepackage{xcolor}
\usepackage{url}
\usepackage{hyperref}
\usepackage{arydshln}
\usepackage{booktabs}
\usepackage{multirow}
\usepackage{xspace}
\usepackage{threeparttable}

\usepackage{tablefootnote}

\usepackage{listings}
\newcommand{\RM}{\textsc{AceCode-RM}\xspace}
\newcommand{\coder}{\textsc{AceCoder}\xspace}
\newcommand{\dataset}{\textsc{AceCode-87K}\xspace}
\definecolor{LightCyan}{rgb}{0.88,1,1}
\newcommand*\samethanks[1][\value{footnote}]{\footnotemark[#1]}

\title{\coder{}: Acing Coder RL via Automated Test-Case Synthesis}

\author{Huaye Zeng$^\dagger$\thanks{Equal Contribution}~, Dongfu Jiang$^\dagger$\samethanks~, Haozhe Wang$^\ddagger$, Ping Nie$^\mathsection$, Xiaotong Chen$^\clubsuit$, Wenhu Chen$^\dagger$\\
    $^\dagger$University of Waterloo, $^\ddagger$HKUST, $^\mathsection$Independent Researcher, $^\clubsuit$Netmind.AI\\
    \texttt{\{w33zeng,dongfu.jiang,wenhuchen\}@uwaterloo.ca} \\
    \\
    \url{https://tiger-ai-lab.github.io/AceCoder}
    }

\begin{document}
\maketitle
\begin{abstract}    
Most progress in recent coder models has been driven by supervised fine-tuning (SFT), while the potential of reinforcement learning (RL) remains largely unexplored, primarily due to the lack of reliable reward data/model in the code domain. In this paper, we address this challenge by leveraging automated large-scale test-case synthesis to enhance code model training. Specifically, we design a pipeline that generates extensive (question, test-cases) pairs from existing code data. Using these test cases, we construct preference pairs based on pass rates over sampled programs to train reward models with Bradley-Terry loss. It shows an average of 10-point improvement for Llama-3.1-8B-Ins and 5-point improvement for Qwen2.5-Coder-7B-Ins through best-of-32 sampling, making the 7B model on par with 236B DeepSeek-V2.5. Furthermore, we conduct reinforcement learning with both reward models and test-case pass rewards, leading to consistent improvements across HumanEval, MBPP, BigCodeBench, and LiveCodeBench (V4). Notably, we follow the R1-style training to start from Qwen2.5-Coder-base directly and show that our RL training can improve model on HumanEval-plus by over 25\% and MBPP-plus by 6\% for merely 80 optimization steps. We believe our results highlight the huge potential of reinforcement learning in coder models. 
\end{abstract}

\section{Introduction}
\label{sec:intro}
In recent years, code generation models have advanced significantly with compute scaling~\citep{kaplan2020scaling} and training data quality improvement~\citep{huang2024opencoder,lozhkov2024starcoder,guo2024deepseek}. The state-of-the-art coder models, including Code-Llama~\citep{codellama}, Qwen2.5-Coder~\citep{Hui2024Qwen25CoderTR}, DeepSeek-Coder~\citep{Guo2024DeepSeekCoderWT} and so on, have shown unprecedented performance across a wide range of coding tasks like program synthesis~\citep{codex}, program repair~\citep{opencodeinterpreter}, optimization~\citep{shypula2023learning}, test generation~\citep{steenhoek2023reinforcement}, SQL~\citep{spider}, and issue fix~\citep{swebench}. These models are all pre-trained and further supervised fine-tuned (SFT) on large-scale coding data from web resources like Common Crawl or Github.

\begin{figure}[!t]
    \centering
    \includegraphics[width=0.5\textwidth]{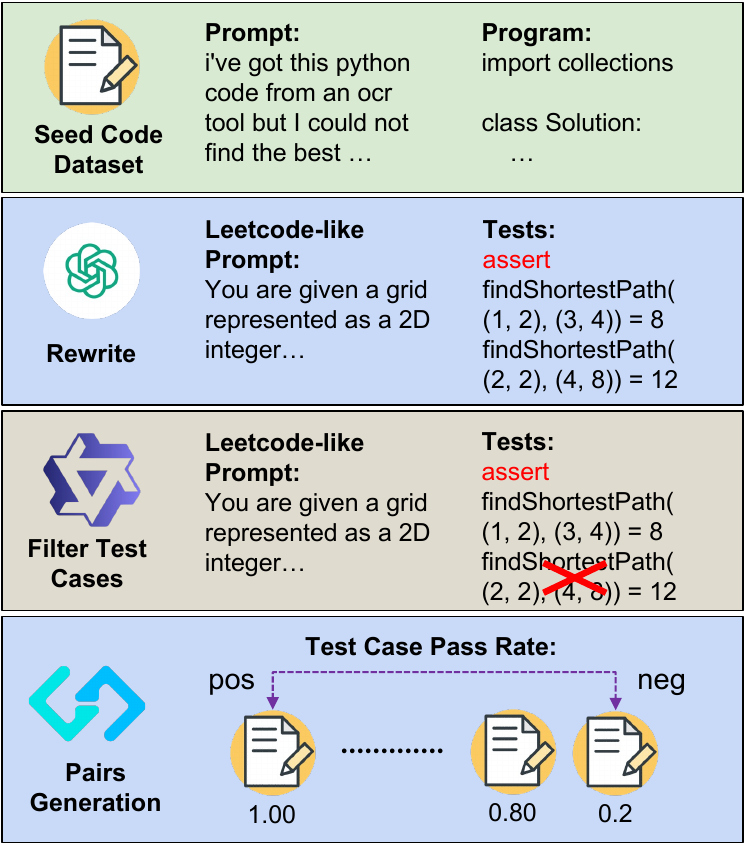}
    \caption{Overall Workflow of our model: we start from the seed code dataset to create well-formatted questions and corresponding test cases. Then we adopt strong models to filter the noisy test cases. Finally, we adopt these test cases to harvest positive and negative program pairs for reward model training and RL.}
    \vspace{-3ex}
    \label{fig:overview}
\end{figure}
Though strong performance has been achieved through SFT~\citep{luo2023wizardcoder,magicoder}, very few models have explored the potential of reinforcement learning (RL)~\citep{Ouyang2022TrainingLM}, which has proven effective in other domains such as mathematical reasoning like DeepSeek-R1~\citep{Shao2024DeepSeekMathPT}. We argue that this absence of RL-based training in coder models is primarily due to two key challenges:\vspace{1ex}\\
\textbf{(1) Lack of reliable reward signals for code generation}. In tasks such as mathematical problem-solving, rewards can be easily derived from rule-based string matches with reference answers~\citep{DeepseekR1} or large-scale human annotations~\citep{instructgpt}. In contrast, evaluating code quality typically requires executing test cases to measure the pass rate, making reward signal design more complex. This also explains why existing reward models like Skywork~\citep{skywork} can hardly generalize to the coding domain (see ~\autoref{subsec:main_results}). \\
\textbf{(2) Scarcity of large-scale coding datasets with reliable test cases.} Most existing coding datasets like APPS~\citep{APPS,codex} heavily rely on costly human expert annotations for test cases, which limits their scalability for training purposes. The largest dataset is TACO~\citep{taco}, containing 25K examples collected from popular coding competition websites, sources that have already been extensively utilized during the pre-training phase.

Therefore, we curate \dataset{}, on which we trained our reward models: \RM-7B and \RM-32B. Comprehensive experiments of best-of-N sampling show that \RM can significantly boost existing LLM's performance on coding benchmarks. For example, \RM-7B can improve the performance of Llama-3.1-8B-Instruct by an average of $8.4$ points across the 4 coding benchmarks, i.e. HumanEval~\citep{evalplus}, MBPP~\citep{evalplus}, BigCodeBench~\citep{zhuo2024bigcodebench} and LiveCodeBench~\citep{jain2024livecodebench}. Even for the stronger coder model Qwen2.5-Coder-7B-Instruct, our "7B+7B" combination still gets an average of $2.6$ improvements. \RM-32B is even more powerful, which pushes the former two numbers to $10.7$ and $4.7$ respectively, showcasing the effectiveness of \RM.

Additionally, both \RM-7B and \RM-32B models demonstrate strong performance on the RM Bench \citep{liu2024rmbench}, outperforming existing state-of-the-art reward models across various categories. Specifically, \RM-32B achieves the highest average score of 76.1, leading in key categories such as Coding, Chat, Normal, and Hard tasks. Even the smaller \RM-7B model shows notable competitiveness, excelling particularly in Coding and Hard categories.

Furthermore, we adopt \RM-7B and test case pass rate separately to do reinforcement learning with reinforce++~\citep{reinforce++} over coder models. Experiments show $2.1$ and $0.7$ points of average improvement when starting from Qwen2.5-7B-Ins and the Qwen2.5-Coder-7B-Ins respectively, making the latter even more powerful than GPT-4-Turbo on benchmarks like MBPP. Inspired by the recent DeepSeek-R1~\citep{DeepseekR1}, we also perform RL training directly from the Qwen2.5-Coder-7B-base model and saw a surprising 25\% improvement on HumanEval-plus and 6\% improvement on MBPP-plus~\citep{evalplus} with merely 80 optimization steps (48 H100 GPU hours). These improvements are also generalizable to other more difficult benchmarks. 


To our knowledge, this is the first work to perform reward model training and reinforcement learning for code generation using a fully automated pipeline that synthesizes large-scale reliable tests.
We believe our \dataset{} will unlock the potential of RL training for code generation models and help the community to further push the boundaries of LLM's coding abilities.

\section{Methodology}
\label{sec:methodology}

In this section, we will introduce the overall methodology of \coder. We begin with formulations of the problems we are investigating, including reward model training and reinforcement learning for LLMs. Finally, we elaborate on how we synthesize the test cases and construct the \dataset{}.

\subsection{Problem Formulation}
\label{subsec:problem_formulation}
\paragraph{Reward Model Training}
Let $x$ denote the coding question and $\mathbf{y} = \{y_1, \cdots, y_t\}$ denote the program solution, where $y_i$ represents the $i$-th token of the program solution and $(\mathbf{x},\mathbf{y}) \in D$. Assuming $\theta$ represents the parameters of the model, then $n$ responses $(\mathbf{y}^1, ..., \mathbf{y}^n)$ will be sampled from the model $\pi_{\theta}$ given the input $\mathbf{x}$. Let $(s_1, ..., s_n)$ be the target rewards, i.e. the test case pass rates in our scenario, then we define the Bradley-Terry loss~\citep{Bradley1952RankAO} for every pair of responses $\mathbf{y}^i$ and $\mathbf{y}^j$ with scores of $s_i$ and $s_j$ when we are training a reward model $R_\phi$ as follows:
\begin{equation*}
\begin{split}
    &\mathcal{L}_\phi(\mathbf{x},s_i,s_j) \\
    & = \mathbbm{1}[s_i > s_j]\log{\sigma(R_\phi(\mathbf{x}, \mathbf{y}^i) - R_\phi(\mathbf{x}, \mathbf{y}^j))}
\end{split}
\end{equation*}
where $\mathbbm{1}[\cdot]=1$ if the expression inside the brackets is true, otherwise, it's 0. The final loss function for the reward training is:
\begin{equation}
\label{eq:rm_loss}
\mathcal{L}(\phi) = -\frac{1}{n(n-1)}\sum_{i=1}^n\sum_{j=1}^n \mathcal{L}_\phi(\mathbf{x},s_i,s_j)
\end{equation}
That means the reward model is trained to assign higher values to preferred responses and lower values to non-preferred ones, maximizing the difference between these ratings.

\paragraph{Best-of-N Sampling}
After we get the trained reward model $R_\phi$, one way to quickly test the performance of the reward model is Best-of-N sampling, which is usually used as a test-time scaling approach. We will simply select the highest-scored response according to the predicted value of $R_\phi$. That is $\mathbf{y}^* = \arg\max_{\mathbf{y}^i \in {\mathbf{y}^1,...,\mathbf{y}^N}} R_\phi(\mathbf{x},\mathbf{y}^i)$. 

\paragraph{Reinforcement Learning}
We can finally conduct reinforcement learning for the original policy model $\pi_{\theta}$ after we get a well-trained reward model $R_\phi$. Proximal Policy Optimization (PPO) is an actor-critic RL algorithm that is widely used for LLM's RL process. Let $\pi_{\theta_{old}}$ be the reference model and $\pi_{\theta}$ be the current policy model that is iteratively updated during the RL training. We denote $r_t(\theta)$ as the probability ratio of the current policy model over the old policy model on the $t$-th generated token:
\begin{equation}
    r_t(\theta) = \frac{\pi_{\theta}(y_t|\mathbf{x}, \mathbf{y}_{<t})}{\pi_{\theta_{old}}(y_t|\mathbf{x}, \mathbf{y}_{<t})}
\end{equation}
Then the PPO algorithms optimize the LLM by the following surrogate objective:
\begin{equation*}
\begin{split}
    &\mathcal{L}_{PPO}(\theta) = \\
    &-\frac{1}{|\mathbf{y}|}\sum_{t=1}^{|\mathbf{y}|}\min \left [r_t\left (\theta\right )A_t, \text{clip}\left (r_t\left (\theta\right ), 1-\epsilon,1+\epsilon\right )A_t \right ]
\end{split}
\end{equation*}
where $\mathbf{y} \sim \pi_{\theta_{old}}(\cdot|x)$, and $A_t$ is the advantage computed through the Generalized Advantage Estimation (GAE)~\citep{Schulman2015HighDimensionalCC} via the rewards generated by $R_\phi$ and the learned value function $V_\psi$. The PPO training objective will force the policy model $\pi$ to increase the probability of generating tokens with higher $A_t$ and decrease the probability ratio of generating tokens with lower $A_t$ until the clipped bounds $1+\epsilon$ and $1-\epsilon$ are reached, respectively.

However, PPO usually requires training an additional value model $V_\psi$ and thus makes the training inefficient. Recently, there have been some other works like Reinforecement++~\citep{reinforce++} that eliminate the need for a value model but instead compute advantage only using the rewards generated by $R_\phi$ and the KL-divergence of the tokens after the $t$-th token. This makes the RL process more efficient and has also proved to be more stable.

\subsection{\dataset{}}
\label{subsec:dataset}
To be able to train a reward model specifically designed for code generation, the first step is to synthesize reliable test cases for each coding problem and use them as training signals. In this section, we explain the whole procedure of constructing \dataset{} step by step. We show the overall statistics in~\autoref{tab:dataset}. 

\paragraph{Test Case Synthesis from Seed Dataset}
\label{subsecp:test_case_synthesis}
We start from existing coding datasets with provided question $\mathbf{x}$ and corresponding program $\mathbf{y}$. Specifically, we combine Magicoder-Evol-Instruct\footnote{\href{https://huggingface.co/datasets/ise-uiuc/Magicoder-Evol-Instruct-110K}{ise-uiuc/Magicoder-Evol-Instruct-110K}}, Magicoder-OSS-Instruct-75K\footnote{\href{https://huggingface.co/datasets/ise-uiuc/Magicoder-OSS-Instruct-75K}{ise-uiuc/Magicoder-OSS-Instruct-75K}}, and StackPyFunction\footnote{\href{https://huggingface.co/datasets/bigcode/stack-dedup-python-fns}{bigcode/stack-dedup-python-fns}} as our seed dataset. We only keep the questions written in Python that contain either a function or a class, resulting in a total of 124K entries. We find that these datasets contain highly noisy questions that cannot be easily evaluated using test cases. Therefore, we feed every question-solution pair ($\mathbf{x}$, $\mathbf{y}$) into a GPT-4o-mini~\citep{gpt4o} to propose a refined LeetCode-style question $\mathbf{x_r}$ with highly structured instructions. Meanwhile, we also prompt it to `imagine' around 20 test cases $(t_1, ..., t_m)$ for each refined coding question $\mathbf{x_r}$ based on its understanding of the expected behavior of the desired program. See prompt template used in~\autoref{appendix:seed_code_prompt}. Please note that we do not use the program solution $\mathbf{y}$ from the existing datasets at all in our final curated \dataset{}. These datasets are purely used as seeds to help LLM formulate well-structured coding problems.

\paragraph{Test Case Filtering}
\label{subsecp:test_case_consistency}
These `imagined' test cases generated from the LLM contain severe hallucinations. To filter out those hallucinated test cases, we facilitated a stronger coder model Qwen2.5-Coder-32B-Instruct~\citep{Hui2024Qwen25CoderTR} as a proxy to perform quality control. Specifically, we prompt it for each $\mathbf{x_r}$ to generate a program $\mathbf{y}^{\prime}$ and then run these programs over the test cases to approximate their quality. We removed all test cases $t_i$ where the generated solution program $\mathbf{y}^{\prime}$ could not pass. Furthermore, we removed questions with fewer than 5 tests after filtering, as these questions might be overly ambiguous. With the above filtering, we constructed the \dataset{} with 87.1K distinct coding questions and 1.38M cleaned test cases, as represented by $(\mathbf{x_r}, (t_1,...,t_{m_c}))$, where $m_c$ represents the number of test cases after filtering.

\begin{table}[!htb]
\centering
\small
\resizebox{\linewidth}{!}{%
\begin{tabular}{ccccc}
\toprule
Subset & Evol & OSS & Stack Python & Overall \\
\midrule
\multicolumn{5}{c}{Before Filtering} \\
\midrule
\# Examples & 36,256 & 37,750 & 50,000 & 124,006 \\
\# Avg Test Cases & 19.33 & 17.21 & 18.27 & 18.26 \\
\midrule
\multicolumn{5}{c}{After Filtering} \\
\midrule
\# Examples & 26,920 & 25,862 & 34,367 & 87,149 \\
\# Avg Test Cases & 15.14 & 16.33 & 16.08 & 15.87 \\
\# Pairs & 89,089 & 91,636 & 126,784 & 307,509 \\
\bottomrule
\end{tabular}
}
\vspace{-0.5em}
\caption{Dataset statistics of \dataset{} before and after test-case filtering.}
\label{tab:dataset}
\vspace{-1em}
\end{table}

\paragraph{Human Case Study}
\label{subsecp:human_case_study_acecode_87k}
Finally, as a last verification step, we conducted a human case study by randomly selecting 40 questions, each with 5 randomly sampled test cases, resulting in 200 manually annotated test cases. Encouragingly, only 3 out of these 200 test cases were found invalid upon review, indicating that our filtering methodology effectively mitigates hallucinations and preserves test-case quality. Moreover, Table \ref{tab:pass_rate_tab} presents various pass-rate statistics for \dataset{}, illustrating the overall difficulty and performance trends.

\begin{table}[h]
\centering
\scriptsize
\resizebox{0.75\linewidth}{!}{%
\begin{tabular}{cc}
\toprule
Pass @ 1 & 34.90\% \\
Pass @ 4 & 38.50\% \\
Pass @ 8 & 39.20\% \\
Pass @ 16 & 39.60\% \\
\midrule
Avg Test Case Pass \% & 70.30\% \\
\midrule
\makecell[c]{\% Question Where All 16\\Inferences Pass All Test Cases} & 22.80\% \\
\bottomrule
\end{tabular}
}
\vspace{-0.5em}
\caption{Dataset pass-rate statistics of \dataset{} after test-case filtering using Qwen2.5-Coder-7B-Instruct (16 responses per query with temperature=1.0)}
\label{tab:pass_rate_tab}
\vspace{-1em}
\end{table}


\paragraph{Preference Pairs Construction}
\label{subsecp:pref_pair_construction}
We propose to use the Bradley-Terry loss to train the reward model as defined in~\autoref{eq:rm_loss}. Therefore, we need to construct (question, [positive program, negative program]) data from \dataset{}. Specifically, we sample programs $(\mathbf{y}^1,...,\mathbf{y}^n)$ from existing models (e.g. Llama-3.1~\citep{grattafiori2024llama3herdmodels}) w.r.t $\mathbf{x}_r$ and utilize the test-case pass rate to distinguish positive and negative programs.\footnote{Appendix \ref{subsec:reward_models_breakdown} contains a detailed model breakdown for how different models are used throughout the training and evaluation process of \RM for more clarifications.} Since the pass rate $s_i$ for the sampled program $\mathbf{y}^i$ can be any number between $[0,1]$, a minor difference in pass rate may not represent that one program is more accurate than another. Therefore, instead of using $\mathbbm{1}[s_i>s_j]$ to select the preference pairs, we have thus modified the selection rules to be:
\begin{equation}
\label{eq:pref_pair_sel}
    \mathbbm{1}[s_i>s_j+0.4,s_i>0.8,s_j>0]
\end{equation}
This is to ensure the preferred program has at least a $0.8$ pass rate to make sure it represents a more correct program. Also, we find that many sampled programs with 0 pass rates can be caused by some small syntax errors or some Python packaging errors during evaluation. We chose not to include them as the preference pair to make sure our constructed datasets represent only the preference based on the valid pass rate. We also ensure that the sampled programs all come from the backbone of $R_\phi$, so the reward model is trained in an on-policy way. After that, we train our reward model $R_\phi$ by fully fine-tuning an instruction coding model. Specifically, we extract the last token's final hidden representation and pass it through a linear model head that generates a single scalar output, which is optimized via the loss function defined in \autoref{eq:rm_loss}.

\section{Experiments}
\label{sec:experiments}
\begin{table*}[!t]
\centering
\small
\resizebox{\textwidth}{!}{%
\begin{tabular}{lccccccccccc}
\toprule
\multirow{2}{*}{Mehod} & \multirow{2}{*}{\# N} & \multicolumn{2}{c}{HumanEval} & \multicolumn{2}{c}{MBPP} & \multicolumn{2}{c}{BigCodeBench-C} & \multicolumn{2}{c}{BigCodeBench-I} & LiveCodeBench & \multirow{2}{*}{\textbf{Average}} \\
 &  & - & Plus & - & Plus & Full & Hard & Full & Hard & V4 &  \\
 \midrule
GPT-4o (0806)       & 1 & 92.7  & 87.2 & 87.6 & 72.2 & 58.9 & 36.5 & 48.0 & 25.0 & 43.6 & 61.3  \\
DeepSeek-V2.5       & 1 & 90.2  & 83.5 & 87.6 & 74.1 & 53.2 & 29.1 & 48.9 & 27.0 & 41.8 & 59.5 \\
DeepSeek-V3         & 1 & 91.5  & 86.6 & 87.6 & 73.0 & 62.2 & 39.9 & 50.0 & 27.7 & 63.5 & 64.6 \\
Qwen2.5-Coder-32B   & 1 & 92.1  & 87.2 & 90.5 & 77.0 & 58.0 & 33.8 & 49.0 & 27.7 & 48.3 &  62.6 \\
 \midrule
\multicolumn{12}{c}{Inference Model = Mistral-7B-Instruct-V0.3} \\
 \midrule
Greedy & 1 & 36.6 & 31.1 & 49.5 & 41.3 & 25.9 & 6.1 & 20.1 & 5.4 & 7.3 & 24.8 \\
Average & 64 & 37.1 & 30.8 & 45.1 & 38.0 & 21.7 & 4.2 & 17.6 & 3.0 & 4.0 & 22.4 \\
Oracle & 64 & 87.2 & 78.0 & 83.9 & 73.5 & 68.4 & 37.8 & 58.5 & 31.1 & 24.3 & 60.3 \\
\midrule
\multirow{3}{*}{AceCodeRM-7B} & 16 & 65.9 & 56.7 & 59.3 & \textbf{52.4} & 35.1 & 10.1 & 29.3 & 8.8 & 11.9 & 36.6 \\
 & 32 & 68.3 & 58.5 & 59.8 & 51.6 & 37.4 & 8.8 & 30.7 & 10.8 & 14.6 & 37.8 \\
 & 64 & 71.3 & 61.6 & 59.8 & 51.6 & 39.4 & 6.8 & 31.8 & 9.5 & 15.4 & 38.6 \\
\rowcolor{LightCyan}
 $\Delta$ (RM-greedy) & - & +34.8 & +30.5 & +10.3 & +11.1 & +13.5 & +4.1 & +11.7 & +5.4 & +8.1 & +13.8 \\
\midrule
\multirow{3}{*}{AceCodeRM-32B} & 16 & 68.3 & 61.0 & 58.7 & 49.5 & 37.7 & 11.5 & 30.9 & 10.1 & 12.9 & 37.8 \\
 & 32 & 72.6 & \textbf{65.9} & \textbf{61.6} & 51.6 & 40.5 & 9.5 & 33.9 & 13.5 & 16.1 & 40.6 \\
 & 64 & \textbf{75.0} & 64.6 & 60.6 & 50.0 & \textbf{42.7} & \textbf{15.5} & \textbf{35.6} & \textbf{13.5} & \textbf{17.4} & \textbf{41.7} \\
\rowcolor{LightCyan}
$\Delta$ (RM-greedy) & - & +38.4 & +34.8 & +12.2 & +11.1 & +16.8 & +9.5 & +15.5 & +8.1 & +10.1 & +16.9 \\
\midrule
\multicolumn{12}{c}{Inference Model = Llama-3.1-8B-Instruct} \\
 \midrule
Greedy & 1 & 68.9 & 62.2 & 67.2 & 54.8 & 38.5 & 12.8 & 31.8 & 13.5 & 18.0 & 40.9 \\
Average & 64 & 61.7 & 54.9 & 64.5 & 54.5 & 32.8 & 10.1 & 26.6 & 9.0 & 13.8 & 36.4 \\
Oracle & 64 & 93.9 & 90.2 & 92.1 & 82.3 & 80.0 & 54.7 & 67.9 & 48.6 & 40.8 & 72.3 \\
\midrule  
\multirow{3}{*}{AceCodeRM-7B} & 16 & 77.4 & 70.7 & \textbf{76.5} & \textbf{64.3} & 45.8 & 20.3 & 36.4 & 12.2 & 26.1 & 47.7 \\
 & 32 & 79.9 & 72.6 & 76.2 & 62.4 & 47.6 & 23.0 & 37.3 & 13.5 & 27.3 & 48.9 \\
 & 64 & 81.7 & 74.4 & 74.6 & 61.9 & 47.8 & \textbf{23.6} & 38.1 & 13.5 & 27.6 & 49.3 \\
\rowcolor{LightCyan}
 $\Delta$ (RM-greedy) & - & +12.8 & +12.2 & +9.3 & +9.5 & +9.3 & +10.8 & +6.2 & 0.0 & +9.6 & +8.4 \\
 \midrule
\multirow{3}{*}{AceCodeRM-32B} & 16 & 82.3 & 74.4 & 72.8 & 60.6 & 49.8 & 20.3 & 38.4 & 13.5 & 27.5 & 48.8 \\
 & 32 & 81.7 & 76.2 & 72.8 & 60.6 & \textbf{50.4} & 22.3 & 39.1 & 13.5 & 30.3 & 49.6 \\
 & 64 & \textbf{85.4} & \textbf{79.3} & 72.0 & 59.0 & 48.5 & 19.6 & \textbf{40.0} & \textbf{13.5} & \textbf{31.0} & \textbf{49.8} \\
\rowcolor{LightCyan}
$\Delta$ (RM-greedy) & - & +16.5 & +17.1 & +9.3 & +9.5 & +11.8 & +10.8 & +8.2 & +0.0 & +13.0 & +9.0 \\
\midrule
\multicolumn{12}{c}{Inference Model = Qwen2.5-Coder-7B-Instruct} \\
 \midrule
Greedy & 1 & \textbf{91.5} & 86.0 & 82.8 & 71.4 & 49.5 & 19.6 & 41.8 & 20.3 & 34.2 & 55.2 \\
Average & 64 & 86.0 & 80.1 & 77.9 & 65.6 & 45.3 & 18.6 & 37.3 & 16.2 & 31.8 & 51.0 \\
Oracle & 64 & 98.2 & 95.7 & 97.4 & 90.7 & 80.9 & 62.8 & 73.5 & 53.4 & 57.4 & 78.9 \\
\midrule
\multirow{3}{*}{AceCodeRM-7B} & 16 & 90.2 & 82.9 & \textbf{88.6} & 74.9 & 53.8 & 20.9 & 45.0 & 21.6 & 40.1 & 57.6 \\
 & 32 & 90.9 & 86.0 & 87.8 & 74.1 & 53.4 & 25.0 & 43.9 & 19.6 & 39.8 & 57.8 \\
 & 64 & 90.9 & 85.4 & 87.6 & 73.8 & 52.9 & 24.3 & 43.5 & 21.6 & 40.1 & 57.8 \\
\rowcolor{LightCyan}
 $\Delta$ (RM-greedy) & - & -0.6 & 0.0 & +5.8 & +3.4 & +4.3 & +5.4 & +3.2 & +1.4 & +5.9 & +2.6 \\
 \midrule
\multirow{3}{*}{AceCodeRM-32B} & 16 & 90.2 & 86.6 & 88.4 & 74.9 & 53.9 & 25.0 & 45.4 & 19.6 & 44.0 & 58.7 \\
 & 32 & 90.2 & \textbf{86.6} & 88.4 & \textbf{75.4} & \textbf{55.4} & \textbf{29.7} & 45.6 & 21.6 & 43.5 & \textbf{59.6}     \\
 & 64 & 89.6 & 86.0 & 87.8 & 75.1 & 55.0 & 26.4 & \textbf{46.1} & \textbf{22.3} & \textbf{44.5} & 59.2 \\
\rowcolor{LightCyan}
$\Delta$ (RM-greedy) & - & -0.6 & +0.6 & +5.8 & +4.0 & +6.0 & +10.1 & +4.3 & +2.0 & +10.3 & +4.4 \\
\bottomrule
\end{tabular}
}
\vspace{-0.5em}
\caption{\RM's best-of-n results on several benchmarks. Specifically, -C means completion split and -I means instruct split of BigCodeBench. The $\Delta$ might be off by 0.1 due to rounding.}
\label{tab:main_bon_results}
\vspace{-1em}
\end{table*}

\subsection{Reward Model Training Setup}
We mainly use Qwen2.5-Coder-7B-Instruct~\footnote{\href{https://huggingface.co/Qwen/Qwen2.5-Coder-7B-Instruct}{Qwen/Qwen2.5-Coder-7B-Instruct}} as the backbone of the reward model and sample 16 responses from it for each question in \dataset{}. Finally, following the rule defined in ~\autoref{eq:pref_pair_sel}, around 300K preference pairs were created out of 46,618 distinct questions (37.34\% of the total questions) that have at least one pair satisfying the condition, and other questions are not used.
\begin{table*}[!t]
\centering
\small
\resizebox{\textwidth}{!}{%
\begin{tabular}{lcccccccccc}
\toprule
\multirow{2}{*}{Model} & \multicolumn{2}{c}{HumanEval} & \multicolumn{2}{c}{MBPP} & \multicolumn{2}{c}{BigCodeBench (C)} & \multicolumn{2}{c}{BigCodeBench (I)} & LiveCodeBench & \multirow{2}{*}{Average} \\
 & - & Plus & - & Plus & Full & Hard & Full & Hard & V4 &  \\
\midrule
RLEF-8B             & -     & 67.5 & -    & 57.0 & -    & -    & -    &  -   & -    & -   \\
RLEF-70B            & -     & 78.5 & -    & 67.6 & -    & -    & -    & -    & -    & -    \\
PPOCoder-7B         & 78.7  & -    & 67.0 & -    & -    & -    & -    & -    & -    & -    \\
StepCoder-7B        & 76.8  & -    & 63.8 & -    & -    & -    & -    & -    & -    & -    \\
CodeGemma-7B        & 60.5  & -    & 55.2 & -    & -    & -    & -    & -    & -    & -    \\
DSTC-33B            & 79.9  &  72.0    &  82.5 & 70.4   & 51.6 & 22.3 & 41.0 & 18.2   & -    & -    \\
\midrule
\multicolumn{11}{c}{Baseline = Qwen2.5-7B-Instruct} \\
\midrule
Baseline        & 81.7 & 73.2 & 79.4 & 67.7 & 45.6 & 16.9 & 38.4 & 14.2 & 29.0 & 49.6 \\
AceCoder$_{RM}$ & 83.5 & \textbf{77.4} & \textbf{83.1} & \textbf{71.2} & \textbf{46.8} & \textbf{16.9} & 39.0 & 14.9 & \textbf{30.3} & \textbf{51.5} \\
AceCoder$_{Rule}$ & \textbf{84.1} & 77.4 & 80.2 & 68.3 & 46.8 & 15.5 & \textbf{40.2} & \textbf{15.5} & 30.1 & 50.9 \\
 \rowcolor{LightCyan}
$\Delta$ (RL-baseline) & +2.4 & +4.3 & +3.7 & +3.4 & +1.2 & 0.0 & +1.8 & +1.4 & +1.3 & +2.0 \\
 \midrule
 \multicolumn{11}{c}{Baseline = Qwen2.5-Coder-7B-Base} \\
\midrule
Baseline & 61.6 & 53.0 & 76.9 & 62.9 & 45.8 & 16.2 & 40.2 & 14.2 & \textbf{28.7} & 44.4 \\
AceCoder$_{RM}$ & 83.5 & 75.6 & 80.2 & 67.2 & 41.9 & 14.9 & 36.8 & 16.2 & 25.7 & 49.1 \\
AceCoder$_{Rule}$ & \textbf{84.1} & \textbf{78.0} & \textbf{82.3} & \textbf{69.3} & \textbf{48.6} & \textbf{18.2} & \textbf{43.2} & \textbf{18.2} & 28.5 & \textbf{52.3} \\
 \rowcolor{LightCyan}
$\Delta$ (RL-baseline) & +22.5 & +25.0 & +5.4 & +6.4 & +2.8 & +2.0 & +3.1 & +4.1 & -0.2 & +7.9 \\
 \midrule
 \multicolumn{11}{c}{Baseline = Qwen2.5-Coder-7B-Instruct} \\
\midrule
Baseline & \textbf{91.5} & \textbf{86.0} & 82.8 & 71.4 & 49.5 & 19.6 & 41.8 & \textbf{20.3} & 34.2 & 55.2 \\
AceCoder$_{RM}$ & 89.0 & 84.1 & \textbf{86.0} & \textbf{72.8} & 50.4 & 18.9 & 42.0 & 19.6 & \textbf{35.0} & 55.3 \\
AceCoder$_{Rule}$ & 90.9 & 84.8 & 84.1 & 71.7 & \textbf{50.9} & \textbf{23.0} & \textbf{43.3} & 19.6 & 34.9 & \textbf{55.9} \\
 \rowcolor{LightCyan}
$\Delta$ (RL-baseline) & -0.6 & -1.2 & +3.2 & +1.3 & +1.4 & +3.4 & +1.5 & -0.7 & +0.8 & +0.7 \\
\bottomrule
\end{tabular}
}
\vspace{-0.5em}
\caption{\coder's Performance after RL tuning using Reinforcement++ algorithm. We start with 3 different initial policy models and 2 kinds of reward types, where $RM$ means using our trained \RM and $Rule$ means using the binary pass rate. Results show consistent improvement across various benchmarks.}
\label{tab:rl_results}
\vspace{-1em}
\end{table*}

Our reward model is trained using LlamaFactory~\citep{zheng2024llamafactory}. We apply full fine-tuning with DeepSpeed stage 3. We train for 1 epoch using a cosine learning rate schedule, starting at 1e-5 with a warmup ratio of 0.1 to gradually increase the learning rate in the initial training phase. The training batch size is set to 128. We enable BF16 precision to reduce memory overhead without compromising model fidelity. The training takes 24 hours on 8 x A100 GPUs.

\subsection{Reinforcement Learning Setup}

We perform RL training from three policy models: Qwen2.5-7B-Instruct~\footnote{\href{https://huggingface.co/Qwen/Qwen2.5-7B-Instruct}{Qwen/Qwen2.5-7B-Instruct}}, Qwen2.5-Coder-7B-Base~\footnote{\href{https://huggingface.co/Qwen/Qwen2.5-Coder-7B}{Qwen/Qwen2.5-Coder-7B}}, and Qwen2.5-Coder-7B-Instruct. Two types of reward can be used, i.e. the trained reward model \RM-7B and the rule-based reward, i.e. pass rate over the test cases in \dataset{}. During training, we set the pass rate to be a binary reward, which is 1.0 when all test cases pass, otherwise 0. This is similar to the verifiable reward used in Tulu3~\citep{tulu3} and DeepSeek-R1~\citep{DeepseekR1}. Similar to DeepSeek-R1~\citep{DeepseekR1}, we also experiment with RL from the base model because SFT may cause the search space of the model to be stuck in a local minimum. Since coding is also a highly verifiable task like math, we include the Qwen2.5-Coder-7B-Base in our experiments.

We have trained different policy model backbones with different rewards, resulting in 6 RL models in total. All the RL-tuning is based on \texttt{OpenRLHF}~\citep{Hu2024OpenRLHFAE}. We adopt the Reinforcement++~\citep{reinforce++} algorithm instead of PPO to improve the training efficiency without training the value model. It's also proved to be more stable than PPO and GRPO. We train our model on a subsampled hard version of \dataset{}, where we keep the 25\% of the questions with lower average pass rates and higher variance. This is to ensure the question is hard and that the sampled programs are diverse enough. For the training hyperparameters, we set the rollout batch size to 256, and 8 programs are sampled per question. The training batch size is 128 with a learning rate of 5e-7. All the models are trained for 1 episode and finished in 6 hours on 8 x H100 GPUs.

\subsection{Evaluation Setup}
We evaluate our method on four established code-focused benchmarks: HumanEval(+)~\citep{codex,evalplus}, MBPP(+)~\citep{austin2021program,evalplus}, BigCodeBench~\citep{zhuo2024bigcodebench} and LiveCodeBench (V4)~\citep{jain2024livecodebench}. 
These benchmarks collectively cover a diverse array of coding tasks, enabling us to assess both the correctness and quality of generated code. For Best-of-N sampling, we adopt top-p sampling with a temperature of 1.0 to generate multiple (16/32/64) candidate solutions per question and then select the response with the highest reward for evaluation. For RL experiments, we use each benchmark's default setting, which is greedy sampling most of the time.

\subsection{Main Results}
\label{subsec:main_results}

Here we show the experimental results of the reward models and RL-trained models.
\paragraph{RM Results}
We conduct Best-of-N experiments on 3 inference models, specifically Mistral-Instruct-V0.3-7B\citep{mistral7binstructv03}, Llama-3.1-Instruct-8B~\citep{grattafiori2024llama3herdmodels}, and Qwen2.5-Coder-7B-Insutrct~\citep{hui2024qwen2, qwen2}. We additionally report the average score across all generated samples and also the oracle score (pass@N) for better comparison.

According to ~\autoref{tab:main_bon_results}, \RM can consistently boost the performance of inference models by a large margin compared to the greedy decoding results. On weaker models like Mistral~\citep{mistral7binstructv03} and Llama-3.1~\citep{zheng2024llamafactory}, the overall improvements are greater than 10 points. These improvements can be attributed to our reward model’s ability to identify high-quality completions among multiple candidates, thereby reducing the impact of suboptimal sampling on the final output. Notably, these gains become more pronounced on benchmarks where the gap between greedy decoding and oracle performance (i.e., the best possible completion among all samples) is larger. In such cases, the variance among sampled completions is relatively high, providing greater opportunities for the reward model to pinpoint and elevate top-tier responses.

Greedy decoding systematically outperforms the average sampled performance, reflecting the strong code generation capability of these inference models. Consequently, while most reward models achieve best-of-N results above the average, we consider a reward model effective only if it surpasses the performance of greedy decoding.

\paragraph{RL Results}
We perform RL training over 3 different initial policy models in~\autoref{tab:rl_results} with model-based and rule-based rewards. When starting from Qwen2.5-Instruct-7B, we can see that RL tuning can consistently improve performance, especially for HumanEval and MBPP. Even for the Plus version with more and harder test cases, the RL-tuned model also has more than $3$ points of improvement. 

When starting from the Qwen2.5-Coder-Instruct-7B itself, we can still observe improvements, especially when using the rule-based reward. For example, we get more than $3.4$ improvement on BigCodeBench-Full-Hard. Using the reward model for RL can also bring a $3.2$ improvement on MBPP. This highlights the charm of self-improvement given the reward model backbone is the same as the initial policy model. We compare our method with other RL-based models like RLEF~\citep{rlef}, PPOCoder~\citep{PPOCoder}, StepCoder~\citep{stepcoder}, DSTC~\citep{dstc}, etc. We show that our 7B model is able to beat these competitors across the evaluation benchmarks.

Another experiment we conduct is to perform RL training directly from the base model Qwen2.5-Coder-7B-base. We show significant improvement, especially through test-case pass rewards on HumanEval, MBPP, and BigCodeBench-I. These results are achieved by only training for 80 steps. We believe further scaling up the training will lead to much larger gains.

\paragraph{Comparison with Other RMs}
\begin{table*}[!t]
\centering
\small
\resizebox{\textwidth}{!}{%
\begin{tabular}{lcccccccccc}
\toprule
Method \& RM & \multicolumn{2}{c}{HumanEval} & \multicolumn{2}{c}{MBPP} & \multicolumn{2}{c}{BigCodeBench-C} & \multicolumn{2}{c}{BigCodeBench-I} & LiveCodeBench & Average \\
 & - & Plus & - & Plus & Full & Hard & Full & Hard & V4 & \\
 \midrule
Greedy & 68.9 & 62.2 & 67.2 & 54.8 & 38.5 & 12.8 & 31.8 & \textbf{13.5} & 18.0 & 40.9 \\
Average & 50.1 & 42.2 & 57.9 & 47.2 & 22.0 & 10.6 & 18.2 & 12.0 & 14.9 & 30.6 \\
\midrule
InternLM2-RM-8B & 57.9 & 55.5 & 66.7 & 54.0 & 38.7 & 8.8 & 29.8 & 8.8 & 15.1 & 37.3 \\
Skywork-Gemma-27B & 73.8 & 67.1 & 64.3 & 53.4 & 40.1 & 14.9 & 32.5 & 12.8 & 23.6 & 42.5 \\
Skywork-Llama-3.1-8B & 67.7 & 61.6 & 69.6 & 56.9 & 40.6 & 10.8 & 31.8 & 12.2 & 18.8 & 41.1 \\
\rowcolor{LightCyan}
$\Delta$ (max(other RM)-greedy) & +4.9 & +4.9 & +2.4 & +2.1 & +2.1 & +2.0 & +0.6 & -0.7 & +5.6 & +2.6 \\
\midrule
\RM-7B & \textbf{77.4} & \textbf{70.7} & \textbf{76.5} & \textbf{64.3} & \textbf{45.8} & \textbf{20.3} & \textbf{36.4} & 12.2 & \textbf{26.1} & \textbf{47.7} \\
\rowcolor{LightCyan}
$\Delta$ (RM-greedy) & +8.5 & +8.5 & +9.3 & +9.5 & +7.3 & +7.4 & +4.6 & -1.4 & +8.1 & +6.8 \\
\bottomrule
\end{tabular}
}
\caption{\RM's performance against other open-sourced reward models in terms of Best-of-16 sampling for Llama-3.1-8B-Inst. We can see the top-ranked RM on Reward Bench get little improvements compared to ours.}
\label{tab:comparison_with_other_rm}
\vspace{-1em}
\end{table*}
We compare our \RM with 3 top-ranked RM on the RewardBench, including InternLM2-RM-8B~\citep{cai2024internlm2}, Skywork-Llama-3.1-8B, and Skywork-Gemma-27B~\citep{skywork}, where results are reported in~\autoref{tab:comparison_with_other_rm}. We can see that these general-purpose RM can hardly improve and sometimes decrease the performance through Best-of-N sampling compared to greedy sampling, showcasing the incapability in identifying the correct generated programs. On the other hand, our \RM surpasses all other publicly released reward models in our evaluation and consistently gets positive gains. These findings further underscore our assumption that previous RM training lacks reliable signals for codes and prove that our RMs can generate reliable and state-of-the-art reward signals in code generation tasks.

\begin{table*}[!htb]
    \centering
    \small
    \resizebox{\textwidth}{!}{%
\begin{threeparttable}
\begin{tabular}{ccccccccc}
\toprule
Model & Code & Chat & Math & Safety & Easy & Normal & Hard & Avg \\
\midrule
Skywork/Skywork-Reward-Llama-3.1-8B & 54.5 & 69.5 & 60.6 & \textbf{95.7} & \textbf{89} & 74.7 & 46.6 & 70.1 \\
NVIDIA/Nemotron-340B-Reward & 59.4 & 71.2 & 59.8 & 87.5 & 81 & 71.4 & 56.1 & 69.5 \\
internlm/internlm2-20b-reward & 56.7 & 63.1 & 66.8 & 86.5 & 82.6 & 71.6 & 50.7 & 68.3 \\
internlm/internlm2-7b-reward & 49.7 & 61.7 & \textbf{71.4} & 85.5 & 85.4 & 70.7 & 45.1 & 67.1 \\
Skywork-Reward-Llama-3.1-8B-v0.2\tnote{1} & 53.4 & 69.2 & 62.1 & 96 & 88.5 & 74 & 47.9 & 70.1 \\

Skywork-Reward-Gemma-2-27B-v0.2\tnote{1} & 45.8 & 49.4 & 50.7 & 48.2 & 50.3 & 48.2 & 47 & 48.5 \\
\midrule
AceCodeRM-7B & 66.9 & 66.7 & 65.3 & 89.9 & 79.9 & 74.4 & 62.2 & 72.2 \\
AceCodeRM-32B & \textbf{72.1} & \textbf{73.7} & 70.5 & 88 & 84.5 & \textbf{78.3} & \textbf{65.5} & \textbf{76.1} \\
\midrule
\rowcolor{LightCyan}
$\Delta$ (AceCodeRM-7B - max(other RM)) & 7.5 & -4.5 & -6.1 & -6.1 & -9.1 & -0.3 & 6.1 & 2.1 \\
\rowcolor{LightCyan}
$\Delta$ (AceCodeRM-32B - max(other RM)) & 12.7 & 2.5 & -0.9 & -8 & -4.5 & 3.6 & 9.4 & 6 \\
\bottomrule
\end{tabular}
\begin{tablenotes}
  \item[1] There is no official result for this model; however, the authors made
        best efforts to extend the original RM Bench code base to adapt to this
        new model.
\end{tablenotes}
\end{threeparttable}
}
\vspace{-0.5em}
\caption{\RM's Performance on RM Bench against various other reward models. We can see that \RM-32B model performed best in Coding, Chat, Normal, Hard, and Average scores against all other reward models.}
\label{tab:rm-bench-result}
\vspace{-1em}
\end{table*}
Moreover, we have also evaluated our \RM against other reward models on RM-Bench in~\autoref{tab:rm-bench-result}, a benchmark designed to assess the reward model's capabilities across code synthesis, mathematical reasoning, and other tasks \citep{liu2024rmbench}. In this head-to-head comparison, \RM emerges as the clear state-of-the-art in coding, hard tasks, and overall average. Remarkably, our 7 billion-parameter variant, \RM-7B, outperforms NVIDIA-Nemotron-340B-Reward\citep{nvidia2024nemotron4340btechnicalreport} by 7.50 points on the coding benchmark, proving that a more compact model can deliver superior reward estimates for code generation. Beyond code, \RM also generalizes well: it beats all other reward models on the average score, underscoring its robust reasoning and dialogue capabilities.

\subsection{Ablation Studies}

\paragraph{Test Case Quality Matters}
\begin{table*}[!tbh]
    \centering
    
    \small
    \resizebox{\textwidth}{!}{%
        \begin{tabular}{lcccccccccc}
        \toprule
        \multirow{2}{*}{Method} & \multicolumn{2}{c}{HumanEval} & \multicolumn{2}{c}{MBPP} & \multicolumn{2}{c}{BigCodeBench-C} & \multicolumn{2}{c}{BigCodeBench-I} & LiveCodeBench & Average \\
         & - & Plus & - & Plus & Full & Hard & Full & Hard & V4 & \\
        \midrule
        \multicolumn{10}{c}{Inference Model = Llama-3.1-8B-Instruct} \\
        \midrule
        RM w/o Test Case Filter & 73.8 & 65.9 & 73.3 & 61.4 & 44.6 & 17.6 & 35.5 & 9.5 & 25.1 & 45.2 \\
        RM w/ Test Filter & \textbf{77.4} & \textbf{70.7} & \textbf{76.5} & \textbf{64.3} & \textbf{45.8} & \textbf{20.3} & \textbf{36.4} & \textbf{12.2} & \textbf{26.1} & \textbf{47.7} \\
        \rowcolor{LightCyan}
        $\Delta$ (w/ Filter - w/o Filter) & +3.7 & +4.9 & +3.2 & +2.9 & +1.2 & +2.7 & +0.9 & +2.7 & +1.0 & +2.5 \\
        \midrule
        \multicolumn{10}{c}{Inference Model = Qwen2.5-Coder-7B-Instruct} \\
        \midrule
        RM w/o Test Case Filter & \textbf{91.5} & \textbf{86.0} & 86.0 & 72.2 & 52.5 & \textbf{21.6} & 43.4 & 19.6 & 36.9 & 56.6 \\
        RM w/ Test Filter & 90.2 & 82.9 & \textbf{88.6} & \textbf{74.9} & \textbf{53.8} & 20.9 & \textbf{45.0} & \textbf{21.6} & \textbf{40.1} & \textbf{57.6} \\
         \rowcolor{LightCyan}
        $\Delta$ (w/ Filter - w/o Filter) & -1.2 & -3.0 & +2.6 & +2.6 & +1.3 & -0.7 & +1.6 & +2.0 & +3.2 & +1.0 \\
        \bottomrule
        \end{tabular}
    }
\vspace{-0.5em}
    \caption{Ablation study on test-case filtering. Results are Best-of-16 sampling performance.}
    \label{tab:consistency_ablation}
\vspace{-0.5em}
\end{table*}

\begin{table*}[!tbh]
    \centering
    
    \small
\resizebox{\textwidth}{!}{%
\begin{tabular}{lcccccccccc}
\toprule
\multirow{2}{*}{Method} & \multicolumn{2}{c}{HumanEval} & \multicolumn{2}{c}{MBPP} & \multicolumn{2}{c}{BigCodeBench-C} & \multicolumn{2}{c}{BigCodeBench-I} & LiveCodeBench & Average \\
 & - & Plus & - & Plus & Full & Hard & Full & Hard & V4 \\
\midrule
\multicolumn{11}{c}{Inference Model = Llama-3.1-8B-Instruct} \\
\midrule
\RM (LLama) & 65.9 & 59.1 & 69.6 & 57.9 & 42.7 & 12.8 & 32.9 & \textbf{13.5} & 19.9 & 41.6 \\
\RM (Qwen) & \textbf{77.4} & \textbf{70.7} & \textbf{76.5} & \textbf{64.3} & \textbf{45.8} & \textbf{20.3} & \textbf{36.4} & 12.2 & \textbf{26.1} & \textbf{47.7} \\
\rowcolor{LightCyan}
$\Delta$ (Qwen-Llama) & +11.6 & +11.6 & +6.9 & +6.3 & +3.1 & +7.4 & +3.5 & -1.4 & +6.2 & +6.1 \\
\midrule
\multicolumn{11}{c}{Inference Model = Qwen2.5-Coder-7B-Instruct} \\
\midrule
\RM (LLama) & 87.8 & 81.7 & 82.0 & 67.7 & 50.5 & \textbf{25.0} & 39.0 & 19.6 & 32.4 & 54.0 \\
\RM (Qwen) & \textbf{90.2} & \textbf{82.9} & \textbf{88.6} & \textbf{74.9} & \textbf{53.8} & 20.9 & \textbf{45.0} & \textbf{21.6} & \textbf{40.1} & \textbf{57.6} \\
\rowcolor{LightCyan}
$\Delta$ (Qwen-Llama) & +2.4 & +1.2 & +6.6 & +7.1 & +3.2 & -4.1 & +6.0 & +2.0 & +7.7 & +2.4 \\
\bottomrule
\end{tabular}
    }
\vspace{-0.5em}
    \caption{Comparison of \RM's performance trained on different base model, where \RM (Llama) is based on Llama-3.1-Inst-8B and \RM (Qwen) is based on Qwen-Coder-2.5-7B-Inst. Results are Best-of-16 sampling performance.}
    \label{tab:backbone_ablation}
\vspace{-1em}
\end{table*}
We also conduct experiments to investigate how filtering the test cases with a proxy model can affect the results. As shown in ~\autoref{tab:consistency_ablation}, training RM on data after the filtering improves the performance significantly, especially for those hard code questions like MBPP-Plus and BigCodeBench-Hard (C/I). We believe this is because the test case filtering can ensure the remaining ones are consistent with each other and thus point to the same implicit program, which improves the quality of the rewards.

\paragraph{RM Backbone Matters}
Our results in ~\autoref{tab:backbone_ablation} clearly show that changing the backbone of the reward model from Llama-3.1 to Qwen2.5 can significantly improve the Best-of-16 performance. This is because the Qwen2.5-Coder models have been pre-trained on way more code-related data compared to the Llama-3.1 models, and thus more knowledgeable when tuning it into a reward model.

\paragraph{Does R1-style Tuning Work?}
Inspired by the recent DeepSeek-R1~\citep{DeepseekR1}, we also conduct the RL directly from the base model without any SFT. It turns out we get huge improvements when using rule-based rewards. For example, we get $25.0$ points of improvements on HumanEval-Plus after training only 6 hours from the Base Model, which is way more efficient than large-scale SFT. What's more, the \coder$_{Rule}$ improve the BigCodeBench-Instruct-Full's performance from $40.2$ to $43.2$, nearly the same performance with DeepSeek-R1-Distill-Qwen-32B ($43.9$) which was directly distilled from the DeepSeek-R1 Model. This further consolidates the finding of DeepSeek-Zero. However, we do find that using reward models for RL tuning can lead to worse results. We attribute this to the potential reward hacking during the tuning process.

\section{Related Works}
\label{sec:related_works}

\subsection{Synthesizing Test Cases}
Automatic test generation is a widely used approach for verifying the correctness of LLM-generated programs. Prior work has commonly employed the same LLM that generates the programs to also generate test cases, selecting the most consistent program from multiple sampled outputs in a self-consistency manner~\citep{Chen2022CodeTCG,Huang2023EnhancingLL,Jiao2024PreferenceOF}. However, these generated test cases often suffer from significant hallucinations. To address this issue, Algo~\citep{Zhang2023ALGOSA} introduced the use of an oracle program solution to improve test case quality. While similar in spirit to our test case filtering approach, Algo constructs its oracle solution by exhaustively enumerating all possible combinations of relevant variables, whereas we leverage a stronger coder LLM to generate the oracle solution. Beyond using test cases as verification signals, Clover~\citep{Sun2023CloverCV} enhances program verification by performing consistency checks between code, docstrings, and formal annotations, incorporating formal verification tools alongside LLMs.


\subsection{Reinforcement Learning for LLM}
\label{subsec:rl_for_llm}
Reinforcement Learning from Human Feedback (RLHF)\citep{instructgpt} has been widely adopted to enhance the capabilities of large language models (LLMs) in various tasks, including conversational interactions and mathematical reasoning\citep{yang2024qwen25mathtechnicalreportmathematical}. Reinforcement learning (RL) algorithms such as PPO\citep{Schulman2017ProximalPO}, GRPO\citep{Shao2024DeepSeekMathPT}, and Reinforcement++\citep{reinforce++} have been employed to fine-tune models using reward signals derived from either learned reward models\citep{Shao2024DeepSeekMathPT} or predefined rule-based heuristics~\citep{DeepseekR1, autocode}.

Given that coding is an inherently verifiable task, recent studies have explored RL techniques that leverage direct execution accuracy as a reward signal. PPOCoder~\citep{Shojaee2023ExecutionbasedCG} and CodeRL~\citep{Le2022CodeRLMC} demonstrated the effectiveness of PPO-based RL for coding tasks, while RLEF~\citep{Gehring2024RLEFGC} extended this approach to multi-turn settings by incorporating execution feedback at each step. StepCoder~\citep{Dou2024StepCoderIC} refined the reward mechanism by assigning rewards at a more granular level, considering only successfully executed lines of code. Additionally, DSTC~\citep{Liu2024DSTCDP} explored the application of Direct Preference Optimization (DPO) to code generation by using self-generated test cases and programs.

Despite these advancements, most prior RL-based approaches for coding have been constrained by the use of pre-annotated datasets such as APPS~\citep{APPS}, which consists of only 5,000 examples, with most problems having a single test case. This limited data availability poses challenges to scalable RL training. Furthermore, the potential of reward models for coding remains largely unexplored. In this work, we address these limitations by automatically synthesizing test cases and leveraging trained reward models for reinforcement learning, demonstrating the scalability and effectiveness of our approach.

\section{Conclusion}

We introduced \coder as the first approach to reward model training and RL tuning for code generation using large-scale, reliable test case synthesis. Our data pipeline produces high-quality verifiable code without relying on the most advanced models, enabling effective reward model training and reinforcement learning. Our method significantly improves Best-of-N performance. However, RL training gains are less pronounced, leaving it as a future work to enhance.

\section*{Limitations}
\paragraph{Test Case Synthesis}
Despite our efforts to enhance the reliability of synthesized test cases through prompt engineering and filtering with a reference solution, inaccuracies can still arise. These errors may stem from an incorrect reference solution or test cases that are too simple, failing to capture challenging edge cases. Consequently, passing all test cases does not necessarily guarantee a program's correctness, leading to noise in the reward model training and reinforcement learning (RL) tuning signals. To address this, future work can leverage stronger large language models (LLMs) to synthesize more rigorous test cases, ensuring the inclusion of harder corner cases. Additionally, using more advanced coding LLMs to generate reference solutions could further improve test case filtering, preserving only high-quality examples.

\paragraph{Reinforcement Learning for Coding}
In this paper, we explored RL tuning using three models and two types of rewards: RM-based and rule-based. While significant improvements are observed when tuning Qwen2.5-7B-Instruct and Qwen2.5-Coder-7B-Base, tuning on Qwen2.5-Coder-7B-Instruct exhibited less pronounced gains due to its strong ability originally. This suggests that the current reward signals may still contain noise. Furthermore, there remains considerable room for improvement, particularly in tuning the Qwen2.5-Coder-7B-Base. Given recent advancements in models such as DeepSeek-R1, future work could further refine RL tuning strategies to achieve better performance with a more fine-grained reward design.

\section*{Ethical Statements}
This work fully complies with the ACL Ethics Policy. We declare that there are no ethical issues in this paper, to the best of our knowledge.

\bibliography{acl_latex}

\clearpage
\appendix
\onecolumn
\section{Appendix}
\label{sec:appendix}
\subsection{More related works}

\paragraph{LLM for Code Generation}
\label{subsec:llm_for_coding}
Large language models (LLMs) have demonstrated significant potential in code generation. Due to the unique nature of coding tasks, specialized coding models such as Code Llama~\citep{codellama} and Qwen Coder~\citep{hui2024qwen2,qwen2} were developed shortly after the emergence of general-purpose LLMs. These models typically undergo a two-phase training process: pre-training and fine-tuning. During pre-training, they are exposed to extensive coding corpora sourced from various internet platforms, including raw text, GitHub repositories, and pull requests. This is followed by supervised fine-tuning, which enhances their instruction-following capabilities.
To assess the performance of these models in code generation, several benchmarks have been established, including MBPP~\citep{austin2021program}, HumanEval~\citep{codex}, EvalPlus~\citep{evalplus, evalperf}, Big Code Bench~\citep{zhuo2024bigcodebench}, and Live Code Bench~\citep{jain2024livecodebench}. These benchmarks usually include a series of prompts or problems for the LLMs to solve, and they also contain test cases to assess the correctness of the generated code.

\paragraph{Reward Models}
\label{subsec:reward_models}
Reward models play a crucial role in aligning LLMs by assigning scalar values to response pairs based on specific evaluation criteria, such as human preference~\citep{instructgpt} and accuracy~\citep{Zhang2025TheLO}. They are widely used in reinforcement learning with human feedback (RLHF) to refine model behavior and in Best-of-N sampling to enhance test-time performance.
However, while general-purpose reward models are effective for assessing human preference, they often struggle with specialized domains like mathematics and coding due to the complexity of these tasks. For instance, even top-ranked reward models from Reward Bench~\citep{Lambert2024RewardBenchER}, such as Skywork-RM~\citep{skywork}, have difficulty providing reliable rewards for these domains. To address this issue, task-specific reward models have been developed, such as Qwen-2.5-Math-PRM~\citep{Zhang2025TheLO} for mathematical reasoning. However, coding reward models have remained largely absent due to the lack of reliable training signals—an issue that our proposed \RM{} aims to address.

\subsection{\RM Model Breakdown}
\label{subsec:reward_models_breakdown}
In this section, we present the different models involved in the training and evaluation process of \RM in a concise table for further clarification.
\begin{table}[h]
\centering
\small
\resizebox{\linewidth}{!}{%
\begin{tabular}{cc}
\toprule
\multicolumn{2}{c}{AceCodeRM-7B} \\
\midrule
Backbone                                         & Qwen2.5-Coder-7B-Instruct                                          \\
Training Responses (Preference Pairs) Generation & Qwen2.5-Coder-7B-Instruct                                          \\
Inference Model                                  & Any Model (Qwen2.5-Coder-7B-Instruct, Llama-3.1-8B-Instruct, etc.) \\
\midrule
\multicolumn{2}{c}{AceCodeRM-32B} \\
\midrule
Backbone                                         & Qwen2.5-Coder-32B-Instruct                                         \\
Training Responses (Preference Pairs) Generation & Qwen2.5-Coder-32B-Instruct                                         \\
Inference Model & Any Model (Qwen2.5-Coder-7B-Instruct, Llama-3.1-8B-Instruct, etc.) \\
\bottomrule
\end{tabular}
}
\vspace{-0.5em}
\caption{Models used during the training and evaluation process of \RM.}
\label{tab:rm_model_clarification}
\vspace{-1em}
\end{table}

\newpage

\subsection{Prompt Template}
\label{appendix:seed_code_prompt}
\begin{table}[!ht]
\centering
\begin{tabular}{|p{0.9\linewidth}|}
\hline
\textbf{system:}\\
You are an AI assistant that helps people with python coding tasks.
\\ \hline

\textbf{user:}\\
You are the latest and best bot aimed at transforming some code snippet into a leetcode style question. You will be provided with a prompt for writing code, along with a reference program that answers the question. Please complete the following for me:\\
1. Come up with a leetcode style question which consists of a well-defined problem. The generated question should meet the following criteria:\\
\quad a. The question is clear and understandable, with enough details to describe what the input and output are.\\
\quad b. The question should be solvable by only implementing 1 function instead of multiple functions or a class. Therefore, please avoid questions which require complicated pipelines.\\
\quad c. The question itself should not require any access to external resource or database.\\
\quad d. Feel free to use part of the original question if necessary. Moreover, please do not ask for runtime and space complexity analysis or any test cases in your response. \\
2. Based on the modified question that you generated in part 1, you need to create around 20 test cases for this modified question. Each test case should be independent assert clauses. The parameters and expected output of each test case should all be constants, **without accessing any external resources**. \\
\\
Here is the original question: \\
\{instruction\} \\
\\
Here is the reference program that answers the question: \\
\verb|```|python \\
\{program\} \\
\verb|```| \\
\\
Now give your modified question and generated test cases in the following json format: \\
\{"question": ..., "tests":["assert ...", "assert ..."]\}.
\\ \hline
\end{tabular}
\caption{Prompt Used for Converting Seed Code Dataset into LeetCode-style Questions and Test Cases}
\label{tab:prompt_template_synthesize_w_int}
\end{table}

\begin{table}[!h]
\centering
\begin{tabular}{|p{0.9\linewidth}|}
\hline
\textbf{system:}\\
You are an AI assistant that helps people with python coding tasks.
\\ \hline

\textbf{user:}\\

You are the latest and best bot aimed at transforming some code snippet into a leetcode style question. You will be provided with a reference program. Please complete the following for me:\\
1. Come up with a leetcode style question which consists of a well-defined problem. The generated question should meet the following criteria:\\
\quad a. The question is clear and understandable, with enough details to describe what the input and output are.\\
\quad b. The question should be solvable by only implementing 1 function instead of multiple functions or a class. Therefore, please avoid questions which require complicated pipelines.\\
\quad c. The question itself should not require any access to external resource or database.\\
\quad d. Feel free to use part of the original question if necessary. Moreover, please do not ask for runtime and space complexity analysis or any test cases in your response. \\
2. Based on the modified question that you generated in part 1, you need to create around 20 test cases for this modified question. Each test case should be independent assert clauses. The parameters and expected output of each test case should all be constants, **without accessing any external resources**.\\
\\
Here is the reference program:\\
\verb|```|python\\
\{program\}\\
\verb|```|\\
\\ \hline
\end{tabular}
\caption{Prompt Used for Converting Seed Code Dataset using only the reference program without instruction into LeetCode-style Questions and Test Cases}
\label{tab:prompt_template_synthesize_wo_inst}
\end{table}

\clearpage

\subsection{Case Studies on HumanEval}
\label{subsec:he_case_studies}
\begin{figure*}[!ht]
    \centering
    \includegraphics[width=1\textwidth]{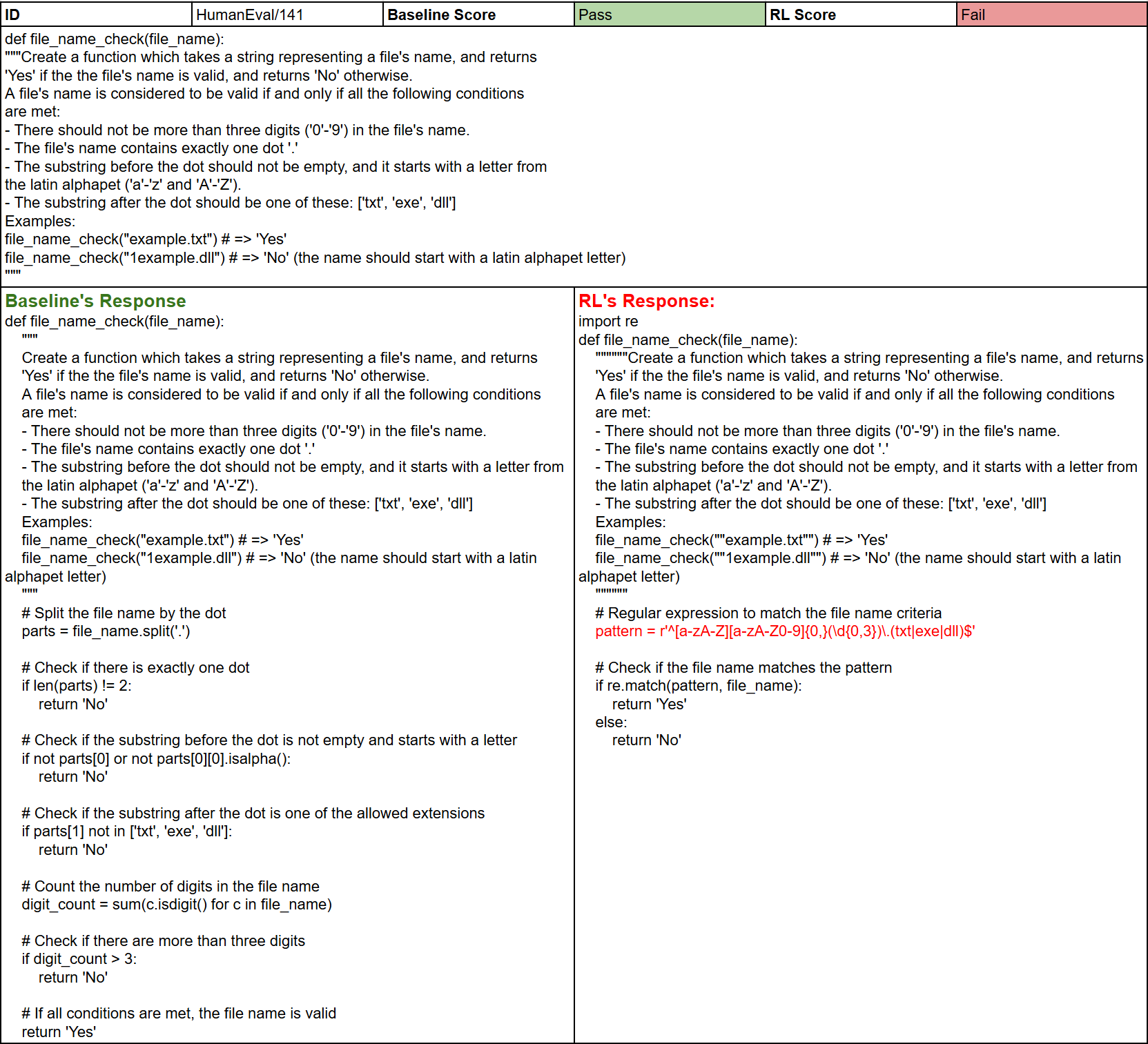}
    \caption{In this example, the RL model took a more advanced approach and attempted to use regular expression matching. However, the regular expression it came up is not correct as it did not include whitespace and did not handle the constraint "there should not be more than three digits ('0'-'9') in the file's name" correctly.}
    \vspace{-3ex}
    \label{fig:he_case_studies_1}
\end{figure*}

\begin{figure*}[!ht]
    \centering
    \includegraphics[width=1\textwidth]{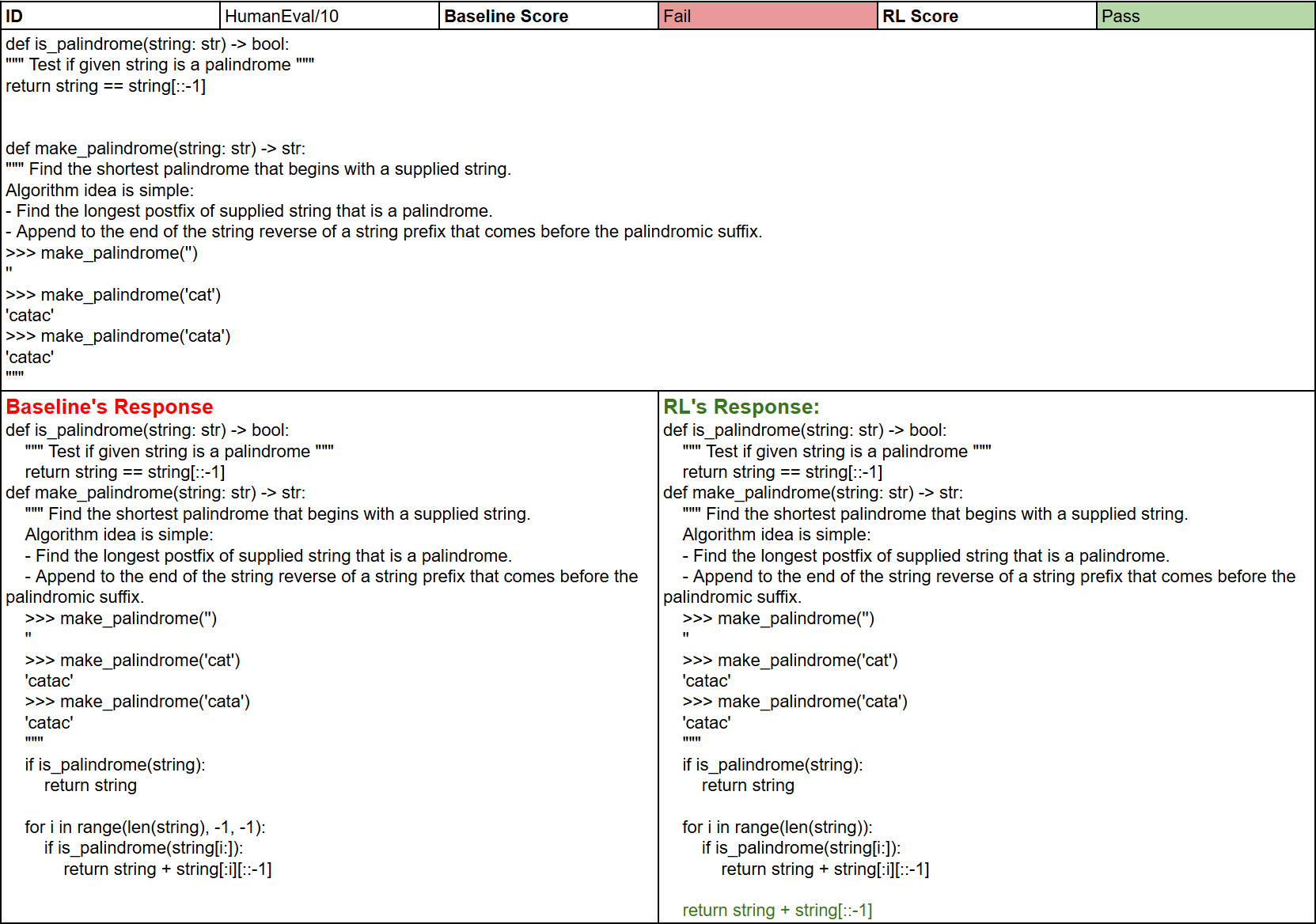}
    \caption{In this example, the RL model correctly caught the error where the baseline did not consider the case where the whole string has to be repeated in order to create a palindrome.}
    \vspace{-3ex}
    \label{fig:he_case_studies_2}
\end{figure*}

\begin{figure*}[!ht]
    \centering
    \includegraphics[width=1\textwidth]{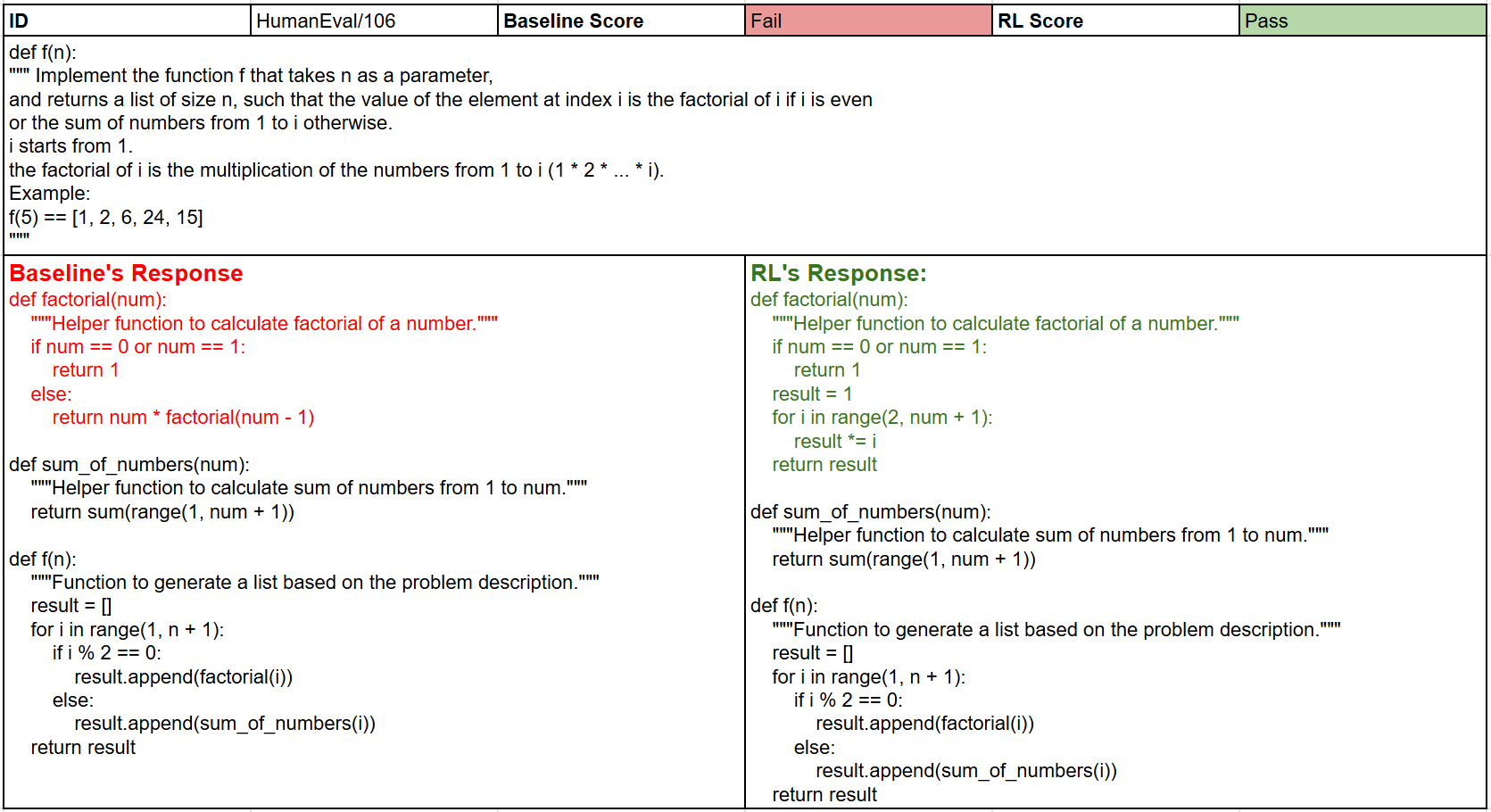}
    \caption{In this example, while both codes would have identical output, the baseline's output is slow due to the recursive calls. By using a for loop instead, the RL model's code's runtime is half of that of the baseline's. Therefore, it passed the test whereas the baseline's code got a time-out.}
    \vspace{-3ex}
    \label{fig:he_case_studies_3}
\end{figure*}

\end{document}